\documentclass[preprint,12pt]{elsarticle}
\usepackage{graphicx,amssymb,latexsym,amsmath,amsbsy}

\newcommand{\ASpace}[1]{{\cal A}_{#1}}             % Object space
\newcommand{\GSpace}[2]{{\cal G}_{#1}^{#2}}        % Data space
\newcommand{\Ntau}{N_{\tau}}                          % number of known values of propagator in imaginary time
\newcommand{\rcut}{M_{\text{cut}}}                     % cut-off for TSVD method
\newcommand{\Eq}[1]{Eq.~(\ref{#1})}                    % Eq. (?)
\newcommand{\Fig}[1]{Fig.~\ref{#1}}                    % Fig. (?)
\newcommand{\apriori}{\textit{a~priori} }              % a priori
\newcommand{\cost}[1]{\mathcal{C}[#1]}                 % cost functional
\newcommand{\Tmax}{\tilde{T}^{*}}                      % maximal value o temperature where gap is visible
\newcommand{\etal}{{\it et al.,\;}}
\newcommand{\beq}{\begin{equation}}
\newcommand{\eeq}{\end{equation}}
\newcommand{\bea}{\begin{eqnarray}}
\newcommand{\eea}{\end{eqnarray}}

\newcommand{\benn}{\begin{displaymath}}
\newcommand{\eenn}{\end{displaymath}}
\journal{Computer Physics Communication}
\begin{document}

\begin{frontmatter}



\title{LINPRO: linear inverse problem library for data contaminated by statistical noise\tnoteref{label1}}
\tnotetext[label1]{This paper and its associated computer program are available via the Computer Physics Communications homepage on ScienceDirect}
 

\author{Piotr Magierski, Gabriel Wlaz\l{}owski}
\address{Faculty of Physics, Warsaw University of Technology,
ulica Koszykowa 75, 00-662 Warsaw, POLAND }
\cortext[cor1]{}

\begin{abstract}
The library \verb|LINPRO| which provides the solution to the linear inverse problem for data
contaminated by a statistical noise is presented. The library makes use of two methods:
Maximum Entropy Method and Singular Value Decomposition.
As an example it has been applied to perform an analytic continuation of the imaginary 
time propagator obtained within the Quantum Monte Carlo method.
\end{abstract}


\end{frontmatter}

\section{Program Summary}
\noindent{\it Title of the program:} 
                  \verb|LINPRO| v1.0

\noindent{\it Catalogue number:}
                  ....

\noindent{\it Program obtainable from:}
                      CPC Program Library, \
                      Queen's University of Belfast, N. Ireland
                      (see application form in this issue)
                      
\noindent{\it Licensing provisions:} GNU Lesser General Public Licence.

\noindent{\it Distribution format:} tar.gz

\noindent{\it Programming language:} C++

\noindent{\it Technical and API documentation:} Yes, in HTML format

\noindent{\it Computer:} \verb|LINPRO| library should compile on any computing system that has C++ compiler.

\noindent{\it Operating systems:} LINUX or UNIX.

\noindent{\it Tested with compilers:} GNU Compiler g++, Intel Compiler icpc.

\noindent{\it External libraries:} OPT++: An Object-Oriented Nonlinear Optimization Library \cite{optpp} (included into distribution).

\noindent{\it No. of lines in distributed program, source files only:} 8 517.

\noindent{\it Nature of problem:} \verb|LINPRO| library solves linear inverse problem with an arbitrary kernel and arbitrary external constraints imposed on the solution.

\noindent{\it Solution method:} \verb|LINPRO| library implements two complementary methods: Maximum Entropy Method and SVD method.

\section{Linear inverse problem}

\subsection{Formulation of the problem}

The inverse problem considered here is of the form:
\beq \label{linp}
G(y) = \int_{-\infty}^{\infty} K(x,y) A(x) dx,
\eeq
where $y\in (\alpha,\beta)$ and the kernel $K$ is a known, real function,
sufficiently regular, although not necessarily smooth.
The function $G$ is known, and is represented by a finite number $\Ntau$
of values at a given set of points: 
$(y_{1}, y_{2}, ..., y_{\Ntau} )$.
The values $G(y_{i})=G_{i}$ and $\vec{G}=(G_{1},G_{2},\ldots,G_{\Ntau})^{T}$ will be called the \textit{data} and
the \textit{data vector}, respectively.
These values are assumed to be in addition affected by a noise
of statistical origin and has to be treated merely as approximations of the true values.
The unknown function $A$ will be called the \textit{object}, irrespective to its physical nature. 
The object is assumed to be nonzero only within a finite interval $(a,b)$, although $a$ and $b$ are in general
unknown. Moreover $A$ may be a subject of additional constraints of the form:
%---------------------------------------------------------------------------
\beq
\int_{-\infty}^{\infty} g_{i}(x) A(x) dx = c_{i}, \hspace{0.5cm} \mbox{$i=1,2,\ldots,L$}
\label{const}
\eeq 
%---------------------------------------------------------------------------
and 
%---------------------------------------------------------------------------
\beq
A(x_{j}) \in [l_{j},u_{j}], \hspace{0.5cm} \mbox{$j=1,2,\ldots,M$},
\label{eqn:const2}
\eeq
%---------------------------------------------------------------------------
where functions $g_{i}$ and values $c_{i}$ are known, $l_{j}$ and $u_{j}$ indicate 
the lower and upper bound imposed on the object at some point $x_{j}$.

\subsection{Normal solution}
Since the function $G$ is known for the finite set of argument values
the linear inverse problem (\ref{linp}) in practice reduces to its
discretized counterpart:
\beq
 G_{i} = \int_{-\infty}^{\infty} K(x,y_{i}) A(x) dx = \int_{-\infty}^{\infty} K_{i}^{*}(x) A(x) dx = (K_{i}, A),
\label{eqn:SVDInverseProblemDefinition}
\eeq
where $(\cdot,\cdot)$ denotes the inner product. The object $A$ can be treated as an element 
of $N$-dimensional Hilbert space $\ASpace{N}$ (in general $N=\infty$). 
Note that due to discretization the kernel functions $K_{i}(x)$ span only $M$-dimensional subspace $\ASpace{M}$ ($M\leq \Ntau$) of the space $\ASpace{N}$. 
It makes the inverse problem \textit{ill-posed}, as there exists an infinite class of solutions 
satisfying \Eq{eqn:SVDInverseProblemDefinition}. Indeed, let us
expand  $K_{i}$ and $A$ in an orthonormal basis $\{u_{k}\}_{k=1}^{N}$ in $\ASpace{N}$:
\begin{eqnarray}
  K_{i}(x) &=&\sum_{k=1}^{M}f_{i k}u_{k}(x),\label{eqn:SVDExpansion1}\\
      A(x) &=&\sum_{k=1}^{N}a_{k}u_{k}(x)=\sum_{k=1}^{M}a_{k}u_{k}(x)+\sum_{k=M+1}^{N}a_{k}u_{k}(x) \nonumber\\
           &=&A_{P}(x)+A_{\perp}(x)\label{eqn:SVDExpansion2},
\end{eqnarray}
where $A_{P}\in\ASpace{M}$ represents the projection of $A$ onto the $M$-dimensional subspace of $\ASpace{N}$ and $A_{\perp}$ is the remaining part, orthogonal to $A_{P}$: $(A_{P},A_{\perp})=0$. Substituting
the above expansions to \Eq{eqn:SVDInverseProblemDefinition} one gets
\begin{equation}
 G_{i}=\sum_{k=1}^{M}\sum_{l=1}^{N}f_{i k}^{*}a_{l}(u_{k},u_{l})=\sum_{k=1}^{M}f_{i k}^{*}a_{k},
\end{equation}
where we have used the property $(u_{k},u_{l})=\delta_{kl}$. 
The last equality shows that $G_{i}$ is independent of $A_{\perp}$, since
 $G_{i}=(K_{i},A)=(K_{i},A_{P})$.
It implies that the data vector $\vec{G}$ allows only for the reconstruction of $A_{P}$. 
The solution $A_{P}$ with the minimal norm is a unique element of the subspace $\ASpace{N}$
and is called the \textit{normal solution} \cite{svd1}.

In the case of data contaminated by a statistical noise the solution of the problem
is also affected by uncertainties. Below we present two strategies which 
allow us to deal with such problems: 
\begin{enumerate}
\item the singular system analysis which uses the Singular Value Decomposition (SVD) to determine $A_{P}$
      and subsequently decrease uncertainties of the normal solution 
      by incorporating constraints imposed on $A$,
\item the Maximum Entropy Method (MEM), which finds the most probable solution,
      under the condition that data represent random numbers normally distributed around the true values,
      and that certain objects $A$ are more probable than the others 
      (so called \apriori information about the object $A$).
\end{enumerate}

\section{Singular system analysis}

\subsection{SVD Method}

The normal solution $A_{P}$ can be determined using the singular value decomposition of the
integral kernel in \Eq{eqn:SVDInverseProblemDefinition} \cite{svd1,svd2,svd3,svd4}. 
Let us rewrite it in the form:
\begin{equation}\label{eqn:SVDOperatorForm} 
  \vec{G} = \mathcal{K} A.
\end{equation} 
The kernel functions $K_{i}(x)$ span $M$-dimensional subspace $\ASpace{M}$ and therefore $\vec{G}$
has only $M$ independent elements \cite{svd1}.
Thus $\vec{G}$ is an element of $M$-dimensional vector space $\GSpace{M}{\Ntau}$. 
$\mathcal{K}$ is an integral operator which transforms an object from $\ASpace{N}$-space into 
a vector of the data space $\GSpace{M}{\Ntau}$. 
The operator can be treated as a rectangular matrix of dimension 
$\Ntau\times N$. We can define also a conjugate operator $\mathcal{K}^{\dagger}$ which transforms vectors from 
$\GSpace{M}{\Ntau}$-space into $\ASpace{N}$-space using the relation:
\begin{equation}
 ( \mathcal{K}u ,\vec{v})=(u, \mathcal{K^{\dagger}}\vec{v} ),\label{eqn:SVDAdjoinOperatorDefinition}
\end{equation}
where $u\in \ASpace{N}$, $\vec{v}\in \GSpace{M}{\Ntau}$ and
the inner product in the data space is defined as follows:
\begin{equation}
(\vec{v},\vec{v'})=\sum_{i=1}^{\Ntau} v_{i} v_{i}',
\end{equation}
which is a useful definition in the case of uncorrelated data\footnote{In the case of correlated data
it is more appropriate to define the inner product as 
$(\vec{v},\vec{v'})=\sum_{i,j=1}^{\Ntau} v_{i}W_{ij}v_{j}$, where the matrix $W$ is the inverse
of the covariance matrix.}.
The conjugate operator $\mathcal{K}^{\dagger}$ can be treated as a rectangular matrix of dimension $N\times\Ntau$. Consequently the operator $\mathcal{K}\mathcal{K}^{\dagger}$ is represented by a square matrix of dimension $\Ntau\times\Ntau$. 
Matrix elements of the operator $\mathcal{K}\mathcal{K}^{\dagger}$ are simply given by
\begin{equation}
 (\mathcal{KK^{\dagger}})_{ij}=(K_{i},K_{j}).
	\label{eqn:SVDOperatorKKStar}
\end{equation} 
Performing the diagonalization of the matrix $\mathcal{KK^{\dagger}}$ enables to determine 
the dimension of the subspace spanned 
by the kernel functions $K_{i}$. Indeed, the operator $\mathcal{K}\mathcal{K}^{\dagger}$ has $M$ positive eigenvalues $\{\lambda_{i}^{2}\}_{i=1}^{M}$, where $M$ is the rank of the operator $\mathcal{K}\mathcal{K}^{\dagger}$. Corresponding eigenvectors $\{\vec{v}_{i}\}_{i=1}^{M}$ form a basis in the data space. Conjugate operator $\mathcal{K}^{\dagger}\mathcal{K}$, which acts in the object space $\ASpace{N}$, among its eigenvalues has the same positive eigenvalues as the operator 
$\mathcal{K}\mathcal{K}^{\dagger}$ and its
eigenfunctions $\{u_{i}\}_{i=1}^{M}$ form the basis of $\ASpace{M}$-space. 
The eigenvalues $\{\lambda_{i}^{2}\}$, the eigenvectors $\{\vec{v}_{i}\}$ and the eigenfunctions $\{u_{i}\}$ form a \textit{singular system} of the operator $\mathcal{K}$ satisfying the 
\textit{shifted eigenvalue problem}:
\begin{equation}\label{eqn:shifted_eigenvalue_problem} 
 \mathcal{K}u_{i}=\lambda_{i}\vec{v}_{i},\qquad \mathcal{K}^{\dagger}\vec{v}_{i}=\lambda_{i}u_{i}.
\end{equation}
The numbers $\lambda_{i}$ are \textit{singular values}, and ${u}_{i}$, $\vec{v}_{i}$ are \textit{singular functions} and \textit{singular vectors}, respectively. The definition of the conjugate operator $\mathcal{K^{\dagger}}$ and equations~(\ref{eqn:shifted_eigenvalue_problem}) 
allow to express the singular functions in the form:
\begin{equation}
 u_{i}(x)=\dfrac{1}{\lambda_{i}}\sum_{k=1}^{\Ntau} K_{k}(x) (\vec{v}_{i})_{k},
\label{eqn:SVDSignularFunctions}
\end{equation}
where $(\vec{v}_{i})_{k}$ denotes k-th element of vector $(\vec{v}_{i})$.

The singular system forms a suitable basis for expansion
of the unknown object $A_{P}$ \cite{svd1}:
\begin{equation}\label{eqn:SVDA_dagger} 
 A_{P}(x)=\sum_{i=1}^{M}b_{i}u_{i}(x),
\end{equation}
where the expansion coefficients are given by
\begin{equation}\label{eqn:SVDcoeff_bi} 
 b_{i}=\dfrac{(\vec{v}_{i},\vec{G})}{\lambda_{i}}.
\end{equation}

%The algorithm of finding the normal solution is as follow:
%\begin{enumerate}
% \item Diagonalize the matrix $\mathcal{KK^{\dagger}}$ given by \Eq{eqn:SVDOperatorKKStar} and find positive %eigenvalues $\{\lambda_{i}^{2}\}_{i=1}^{M}$ and corresponding eigenvectors $\{\vec{v}_{i}\}_{i=1}^{M}$. This %information is sufficient to determination the expansion coefficients $b_{i}$.
%\item Using the \Eq{eqn:SVDSignularFunctions} produce the singular functions $\{u_{i}\}_{i=1}^{M}$.
%\end{enumerate}

\subsection{Data with noise}
The solution given by \Eq{eqn:SVDA_dagger} can be used only in the case of noiseless data. In the case 
when data vector $\vec{G}$ is known with some uncertainty $\Delta\vec{G}$ the above algorithm 
becomes numerically \textit{ill-conditioned} \cite{svd1,svd2,svd3}. 
Note that the singular values $\{\lambda_{i}\}$, the singular vectors $\{\vec{v}_{i}\}$ and therefore the singular functions $\{u_{i}\}$ are known exactly since they are fully determined by the kernel functions $K_{i}$. 
Errors $\Delta\vec{G}$ affect only the expansion coefficients, which will be the subject to some uncertainty 
$\Delta b_{i}=(\vec{v}_{i},\Delta\vec{G})/\lambda_{i}$. To perceive the origin of numerical instabilities 
let us arrange the set of singular values $\{\lambda_{i}\}_{i=1}^{M}$ in descending order: $\lambda_{1}\geqslant\lambda_{2}\geqslant\ldots\geqslant\lambda_{M}$. Clearly, with a
decreasing singular value the contribution of the statistical noise to $A_{P}$ is amplified:
\begin{equation}
 \lambda_{i}\rightarrow 0\quad \Rightarrow \quad 
\Delta b_{i}=\dfrac{(\vec{v}_{i},\Delta\vec{G})}{\lambda_{i}}\rightarrow\infty.
\end{equation}
Practically it means that in the object space $\ASpace{M}$ there exist ``directions'' which are 
\textit{invisible} for the SVD method, namely, the expansion coefficients cannot be determined with 
sufficient accuracy starting from some $i$ index. 
Simultaneously the corresponding singular functions $u_{i}$ become rapidly oscillating
with an increasing index $i$ (number of nodes of $i$-th singular function is $i-1$) \cite{svd2}. The functions 
associated with smaller values of $\lambda_{i}$ are responsible for reconstructing more subtle  
details of the solution. Since
large uncertainties of coefficients in general yield to strong fluctuations of the solution, 
one of the standard methods is to remove all such strongly fluctuating terms 
and include only those for which $b_{i}$ are determined with satisfactory accuracy:
\begin{equation}\label{eqn:SVDA_tilde} 
A_{P_{\textrm{cut}}}(x)=\sum_{i=1}^{\rcut}b_{i}u_{i}(x).
\end{equation}
This approach leads to the so called truncated SVD method (TSVD). In practice the truncation parameter is chosen in such a way to remove all terms for which the ratio (cut-off parameter) 
$\lambda_{i}/\lambda_{1}$ is smaller than 
$\frac{1}{N_{\tau}}\sum_{i=1}^{N_{\tau}}\frac{\Delta G_{i}}{|G_{i}|}$. 
It ensures
that the solution $A_{P_{\textrm{cut}}}$ reproduces data $G_{i}$ within its error bars and prevents the inclusion of unjustified structures into the solution \cite{svd2,svd3}.

\subsection{Incorporating \apriori information}

The reconstruction quality of 
the SVD method decreases significantly if data are affected by even a relatively weak noise.
It turns out however that the incorporation of \apriori information can improve the reconstruction process \cite{svd4,svd5}.
There are two types of the prior information: information concerning the support of the solution (interval where
the solution is nonzero) and external constraints.
 The first type of information leads to the following modification of the original problem:
\begin{equation}
 G_{i}=\int_{-\infty}^{+\infty}K_{i}^{*}(x)A(x)dx\cong\int_{a}^{b}K_{i}^{*}(x)A(x)dx=\int_{-\infty}^{+\infty}K_{i}^{*}(x)S(x,a,b)A(x),
\end{equation}
where object $A$ is assumed to be non zero in the interval $(a,b)$. 
$S(x,a,b)$ denotes the \textit{support function} defined as
\begin{equation}
 S(x,a,b)=\left\lbrace\begin{array}{ll}
   1,& \textrm{if\;}x\in (a,b) \\
   0,& \textrm{if\;}x\notin (a,b)
                           \end{array} \right. , 
\label{eqn:SVDSupportFunction}
\end{equation} 
which modifies the kernel functions for the SVD method. 
This modification has two major consequences. First, the singular values $\lambda_{i}$ decrease 
faster as the size $|b-a|$ of the support function gets smaller. It subsequently leads to smaller values 
of $\rcut$ and in general decreases the reconstruction ability of the method. Second, however, 
the singular functions $u_{i}$
become limited to the interval $x\in (a,b)$ and  their zeros are spaced more closely. This implies 
that a smaller number of singular functions are needed to get the same accuracy of reconstruction as before. It turns out that the latter consequence dominates and a properly chosen support function increases reconstruction quality \cite{svd5}.

Within the SVD method it is also possible to generate the solution which satisfies integral constraints (\ref{const}). 
This can be done using the fact that each solution of the form
\begin{equation}\label{eqn:SVDA_dagger_tilde} 
 \tilde{A}_{P}(x,\{\tilde{b}_{i}\})=\sum_{i=1}^{M}\tilde{b}_{i}u_{i}(x),
\end{equation}
where $\tilde{b}_{i}\in(b_{i}-\Delta b_{i},b_{i}+\Delta b_{i})$ reproduces the data $G_{i}$ within its error bars. 
Hence choosing an appropriate set of the expansion coefficients 
$\{\tilde{b}_{i}\}_{i=1}^{M}$ one can try to reproduce constraints (it is not always possible since 
the normal solution need not fulfill the same constraints as the true solution) \cite{svd4}.
In general, the expansion coefficients $\{\tilde{b}_{i}\}_{i=1}^{M}$ which agree
with the constraints are not unique. To distinguish between various possibilities one can
define the \textit{cost functional} $\cost{\tilde{A}_{P}}$, which has to be minimized to 
find the best set of coefficients. As a cost functional one can use $\chi^{2}$ statistics with a similar form 
like in the maximum entropy method (see next section). Another possibility is to choose the cost functional as the norm of the solution, $\cost{\tilde{A}_{P}}=||\tilde{A}_{P}||=\sqrt{(A_{P},A_{P})}$. This choice is in agreement with the spirit of an SVD approach, where the normal solution is defined as the solution with the minimal norm. Summarizing, the problem of 
determining the unknown object satisfying external constraints has been reduced to the optimization problem:
\begin{equation}
 \tilde{A}_{P}(x)=\min_{\{\tilde{b}_{i}\}} \cost{\tilde{A}_{P}(x,\{\tilde{b}_{i}\})}
\end{equation}
with external constraints:
\begin{eqnarray}
 \forall i=1,2,\ldots, M &:&b_{i}-\Delta b_{i}\leqslant\tilde{b}_{i}\leqslant b_{i}+\Delta b_{i}, \\
 \forall j=1,2,\ldots, L &:& \int_{-\infty}^{+\infty}g_{j}(x)\tilde{A}_{P}(x)dx=c_{j}.
\end{eqnarray}

\section{Maximum Entropy Method}
\subsection{General considerations}

Let us distinguish between the exact values of the function $G$ which
are unknown and fulfill \Eq{linp} and their known approximations contained in a vector:
$\vec{\tilde{G}}=( \tilde{G}_{1}, \tilde{G}_{2}, \ldots, \tilde{G}_{\Ntau} )^{T}$. These values 
can be treated as a particular realization of random variables, which are assumed to be 
uncorrelated\footnote{The extension of the method to the case of correlated data is straightforward,
but require additional information in the form of covariance matrix.}
and have a normal distribution around the exact values $G_{i}$ with a variance $\sigma_{i}^{2}$.
The probability of obtaining the particular realization $\vec{\tilde{G}}$ under the condition
that the exact values are given by $\vec{G}$ reads
\beq \label{prob}
p(\vec{\tilde{G}} | \vec{G} )\propto 
\exp\left ( -\frac{1}{2}\sum_{i=1}^{\Ntau}\left ( \frac{\tilde{G}_{i}-G_{i}}{\sigma_{i}}\right )^{2}\right ),
\eeq 
and the values $G_{i}$ depend on the function $A$ 
according to the relation (\ref{eqn:SVDInverseProblemDefinition}).
This equation is subsequently discretized in a chosen interval $(a,b)$ and becomes 
a linear transformation:
\beq \label{trans}
G_{i}=\sum_{j=1}^{N} K_{ij} A_{j},
\eeq 
where $K_{ij} = K(x_{j},y_{i} )\Delta x$ is a rectangular matrix $\Ntau\times N$, 
$\Delta x = x_{j} - x_{j-1}$ and $A_{j}=A(x_{j})$. Points $x_{j}$ are uniformly distributed over the interval $( a=x_{1},b=x_{N} )$.

The estimator for the quantity $\vec{A}=(A_{1},A_{2},\ldots,A_{N})^{T}$ is defined as the one which maximizes the conditional
probability $p( \vec{A} | \tilde{G} )$. This in turn can be expressed by (\ref{prob}) using Bayes' theorem:
\beq
p( \vec{A} | \vec{\tilde{G}} ) = \frac{ p(\vec{\tilde{G}} | \vec{G} ) p( \vec{A} ) }{p( \vec{\tilde{G}} )},
\eeq
where on the rhs the dependence on $\vec{A}$ is included in $\vec{G}$ through the relation (\ref{trans}).
The probability $p( \vec{A} )$ is {\em a priori} probability and may contain additional information about
$\vec{A}$ including constraints (\ref{const}). The maximization of this probability (so called {\em likelihood function}) leads in practice to the condition:
\beq
\frac{\partial}{\partial A_{j}}p( \vec{A} | \vec{\tilde{G}} ) = 0, \hspace{0.5cm} \mbox{j=1,..,N},
\eeq
In the case of $N\le \Ntau$ and $p( \vec{A} ) = const$ the above condition reduces to the least square problem
with the solution: $ \vec{A} = (K^{T}K)^{-1}\vec{G}^{T}K $, where $\sigma_{i}=\sigma=const$ is assumed.

Here we are interested in the case when $N > \Ntau$ and an additional prior information is needed.
It is specified through the entropy $S(\vec{A})$, where 
$p( \vec{A} ) \propto \exp ( S(\vec{A}) )$. The completely non-informative entropy is of the form:
\beq
S(\vec{A})=-\alpha \sum_{i=1}^{N} \left ( \frac{A_{i}}{\sum_{j=1}^{N}A_{j}} \right )
           \log \left ( \frac{A_{i}}{\sum_{j=1}^{N}A_{j}} \right ),
\eeq
where $\alpha>0$ is arbitrary. 
It favors the solution $\vec{A}=\vec{const}$.
Usually we have additional information about the structure of $A$ which allows us
to specify a model of $\vec{A}$. In such a case the relative entropy can be constructed:
\beq
S(\vec{A}|\vec{\cal M}) = - \alpha \sum_{i=1}^{N}
\frac{A_{i}\log\frac{A_{i}}{{\cal M}_{i}}}{\sum_{j=1}^{N}A_{j}},
\eeq
where $\vec{\cal M}=( {\cal M}_{1},...,{\cal M}_{N} )^{T}, {\cal M}_{i}={\cal M}(x_{i})$ 
is an assumed model for $\vec{A}$ and $\sum_{i=1}^{N}A_{i} = \sum_{i=1}^{N}{\cal M}_{i}$.
The model $\vec{\cal M}$ has to fulfill the constraints imposed on $\vec{A}$.
In order to be able to construct the entropy in the above form, requires the assumption 
of nonnegativity of $A$. 
In order to avoid a complicated notation we assume also that both $\vec{A}$ and $\vec{\cal M}$
are normalized: $\sum_{i=1}^{N}A_{i} = \sum_{i=1}^{N}{\cal M}_{i}=1$.
Clearly the entropy is maximized in the case when $\vec{A}=\vec{\cal M}$, although note that
$S(\vec{A}|\vec{\cal M})\neq S(\vec{{\cal M}}|\vec{A}) $. 

The prior information provides additional conditions for $\vec{A}$ and makes
the maximization of the {\em likelihood function} a well defined process with a unique solution.
Clearly now:
\beq
p( \vec{A} | \tilde{G} )\propto \exp\left ( -\frac{1}{2}\sum_{i=1}^{\Ntau}\left ( \frac{\tilde{G}_{i}-G_{i}}{\sigma_{i}}\right )^{2} - \alpha\sum_{i=1}^{N} A_{i}\log\frac{A_{i}}{{\cal M}_{i}} \right ),
\eeq
and the maximum entropy method leads to the maximization of the above function with respect to $\vec{A}$ \cite{jaynes}.
Note that still one has a freedom of choosing the constant $\alpha>0$. It governs the relative importance
of the two terms in the above expression and larger $\alpha$ favors the model over the data.
% In practice, since $\sigma$ is usually not known, the parameter responsible for the relative importance of the entropy term is $k\sigma$. 

Another extension of the above formulation which will be considered in the next section admits the
possibility of having a class of models $\vec{\cal M}(x;\vec{f})$, where $\vec{f}$ is a set of parameters
describing admissible degrees of freedom of the model and thus defining a set of admissible models.

\subsection{Method of solution}\label{sec:methodmem}

The quantity which has to be minimized as a result of the MEM reads:
\beq
F(\vec{A}) = \frac{1}{2}\sum_{i=1}^{\Ntau}\left ( \frac{\tilde{G}_{i}-G_{i}}{\sigma_{i}}\right )^{2} + 
\alpha\sum_{i=1}^{N} A_{i}\log\frac{A_{i}}{{\cal M}_{i}}  .
\eeq
The task of minimizing the function of $N$ variables, where $N$ in practice may be of the order
of $10^{2-4}$ is rather hard. Therefore we apply here the procedure described in Ref.~\cite{koonin}
which replaces the minimization of the many-variable function by an iterative process of consecutive
least square problems.
Let us assume that $F(\vec{A}^{0}) = min$ and $\vec{A}^{0}$ represents the solution of the problem.
We expand $F$ around $\vec{A}^{0}$ up to the second order:
\bea
& & F(\vec{A}^{0} + \delta\vec{A}) = F(\vec{A}^{0}, \vec{A}) = \nonumber \\
 &=& \frac{1}{2}\sum_{i=1}^{\Ntau}\left ( \frac{\tilde{G}_{i}-G_{i}}{\sigma_{i}}\right )^{2} 
+ \alpha\sum_{i=1}^{N}\left ( \frac{1}{2 A^{0}_{i}} (\gamma_{i} - A_{i})^{2} + \omega_{i} \right ) + O( |\delta\vec{A}|^{3} ),
\eea
where
\bea
\gamma_{i}&=&A^{0}_{i} \left ( 1-\log\frac{A^{0}_{i}}{{\cal M}_{i}} \right ), \nonumber \\
\omega_{i}&=&{\cal M}_{i} - A^{0}_{i} \left ( 1 - \log\frac{A^{0}_{i}}{{\cal M}_{i}} + 
            \frac{1}{2}\left (\log\frac{A^{0}_{i}}{{\cal M}_{i}} \right )^2 \right ), \\
\vec{A} &=& \vec{A}^{0} + \delta\vec{A}. \nonumber
\eea
The above expansion implies the method of solving the problem. Namely,
in the step $n$ we minimize $F(\vec{A}^{0(n)}, \vec{A}^{(n)})$ 
with respect to $\vec{A}^{(n)}$ at fixed $\vec{A}^{0(n)}$. This is equivalent to the least square problem. 
Then we define a new $\vec{A}^{0(n+1)} = \xi \vec{A}^{(n+1)} + (1-\xi)\vec{A}^{(n)}$, where $\xi\in (0,1)$.
Such a procedure leads to a convergent solution providing $\xi$ is sufficiently small.
As a starting condition one takes $\vec{A}^{(0)} = \vec{A}^{(1)} = \vec{\cal M}$.

The additional constraints (\ref{const}) can be included by considering the modified function:
\beq
G(\vec{A}^{0}, \vec{A}) = F(\vec{A}^{0}, \vec{A}) + \sum_{i=1}^{L} \theta_{i} ( c_{i} - \sum_{j=1}^{N}g_{i,j}A_{j} )^{2},
\eeq
where $g_{i,j}=g_{i}(x_{j})\Delta x$, and $\theta_{i}$ are positive parameters 
governing the ''stiffness'' of the constraints and thus 
responsible for the accuracy at which conditions (\ref{const}) are fulfilled. 

In the MEM approach we have improved the method of finding the solution
by constructing a sequence of minimizations with a gradually refined model. 
In this case the model is of the form $\vec{\cal M}(x;\vec{f})$ and thus represents a class
of models defined by parameters $\vec{f}=(f_{1},...,f_{s})$.
At the end
of each minimization process described above the result has been used to define a new model which maximize the overlap with respect to parameters $\vec{f}$.
\beq
O(\vec{f})=\frac{ \left ( \sum_{i=1}^{N} A_{i} {\cal M}_{i}(\vec{f}) \right )^{2} }
                {\sum_{i=1}^{N} A_{i}^{2} \sum_{i=1}^{N} {\cal M}_{i}^{2}(\vec{f}) },
\eeq 
The above quantity is clearly nonnegative and moreover $O(\vec{f})\in [0,1]$. It is equal to unity
if $A_{i} = {\cal M}_{i}$. 
After the maximization of the overlap the new minimization process is started as described above.
The procedure has been
continued until the value of $|\vec{A}^{(n)} - \vec{A}^{(n-1)}| < \epsilon $, with 
an admissible tolerance $\epsilon > 0$.  This strategy will be called as ``self-consistent'' Maximum Entropy Method.

% \subsection{Tests and resolution}
% {\bf to be done}

\section{Structure of the library}
\subsection{General overview}
\verb|LINPRO| is an object-oriented library for solving linear inverse problems written in C++.
As an optimization engine it uses OPT++ 2.4 library \cite{optpp}. 
The aim of the \verb|LINPRO|  library is to collect in one place various algorithms for solving inverse problems
and provide unified and user friendly programming interface to all of them. The library can be used to solve the problem with an arbitrary kernel defined by the user. For the MEM the package provides an interface for defining the arbitrary default model as well as a class of default models parametrized by a  set of parameters.

\subsection{Installation and technical documentation}
To install the library unpack tarball and follow instructions contained in INSTALL file. The distribution contains also folder \verb|doc| where the technical and the API documentation is located. The documentation is generated in the HTML format, the master file is index.html. The user will find the codes which solve the artificial problem (as presented below), in 
the attached folder: \verb|examples|.

\subsection{Inverse problem solvers}
\begin{figure}[h]
 \centering
 \includegraphics[scale=0.85]{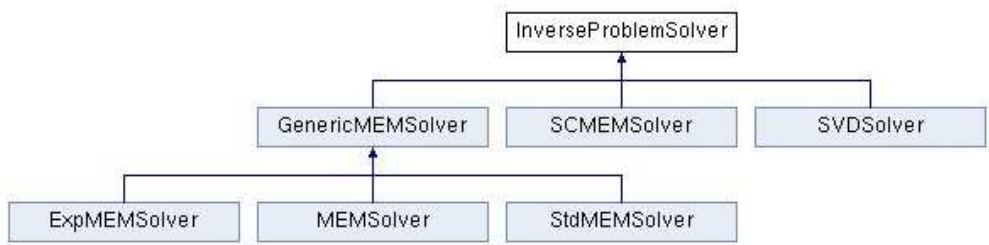}
 \caption{(Color online) Inheritance diagram for InverseProblemSolver.}
 \label{fig:solvers}
\end{figure}

Within \verb|LINPRO| library algorithms for solving the inverse problem are called \textit{solvers}, 
represented by \verb|InverseProblemSolver| class. \Fig{fig:solvers} presents the inheritance diagram of available solvers.

The solvers represents algorithms:
\begin{itemize}
 \item \verb|StdMEMSolver| - standard implementation of the Maximum Entropy Method. To minimize the likelihood function the algorithm reduces the minimization problem to the iterative process of consecutive least square problems, as described in section \ref{sec:methodmem}.
 \item \verb|MEMSolver| - implementation of the Maximum Entropy Method. It minimizes the likelihood function using
 the nonlinear interior-point method.
 \item \verb|ExpMEMSolver| - implementation of the Maximum Entropy Method, where object $A(x)$ is parametrized by the formula $A(x)={\cal M}(x)\exp f(x)$, where ${\cal M}(x)$ is a model function and $f(x)$ is determined by the solver. Such a substitution is often used to eliminate the term $\log A(x)/{\cal M}(x)$, which in specific situations is a source of optimizer instabilities (for example such instabilities can occur if 
 a chosen model ${\cal M}(x)$ is very close to zero for some values of $x$,
 and due to finite precision is treated as zero). To minimize the likelihood function the nonlinear interior-point method is used. Since this solver is much slower than standard solvers it should be used in cases when \verb|StdMEMSolver| and \verb|MEMSolver| do not converge properly.
 \item \verb|SVDSolver| - implementation of the SVD method.
 \item \verb|SCMEMSolver| - implementation of the self-consistent engine for Maximum Entropy Method. It works with each solver belonging to \verb|GenericMEMSolver| branch. The scheme of the solver algorithm presents \Fig{fig:scmemsolver}.
 \begin{figure}[h]
 \centering
 \includegraphics[scale=0.6]{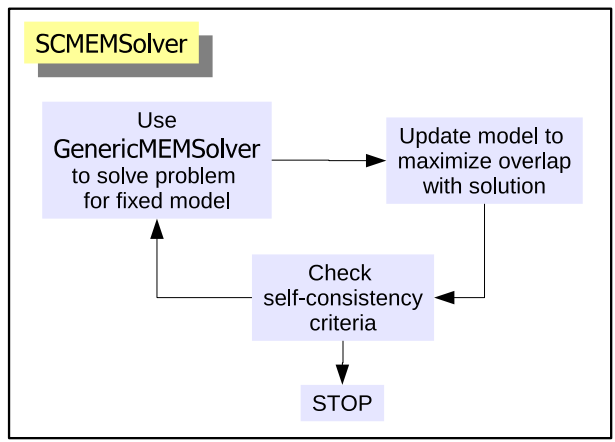}
 \caption{(Color online) The algorithm scheme for SCMEMSolver.}
 \label{fig:scmemsolver}
\end{figure}
\end{itemize}

\section{Example of physical application}

\subsection{Problem formulation}
As an example the package has been applied to extract the spectral weight function $A(x)$
through the analytic continuation of the imaginary time propagator $G(y)$:
\beq
G(y)=-\frac{1}{2\pi}\int_{-\infty}^{\infty}
dx A(x)\frac{\exp(-xy)}{1+\exp(-x\beta)}.
\label{eqn:G_tau}
\eeq
By definition, $A(x)$ fulfills the following constraints:
\bea
& & A(x) \ge 0, \quad \quad
\int_{-\infty}^{+\infty}\frac{dx}{2\pi} A(x) = 1,  \label{eqn:Ap_con1} \\
& & \int_{-\infty}^{+\infty}\frac{dx}{2\pi} A(x)\frac{1}{1+\exp(x\beta)} = -G(\beta).
\label{eqn:Ap_con2}
\eea
This problem is frequently encountered in Quantum Monte Carlo simulations, which by construction
produce data affected by the statistical noise \cite{others,ours}.
In order to check the reconstruction ability of the package an artificially generated data for 
the imaginary time propagator has been used. 
The application of the package to the real physical data can be found in Refs. \cite{ours}. 

The artificial spectral function has been chosen in the form:
\beq
A(x)=\dfrac{1}{2}N(x;-1.5,0.5) + \dfrac{1}{2}N(x;2.0,0.7),
\eeq
where $N(x;\mu,\sigma)$ is the normal distribution function with the mean $\mu$ and 
the standard deviation $\sigma$. Subsequently
the values of the imaginary time propagator has been generated using the relation (\ref{eqn:G_tau})
for $\Ntau$ uniformly spaced data points in the interval $[0,\beta=10]$.

%The codes used to generate results for this problem are attached to the distribution in folder \verb|%examples/tests|.
\subsection{Data without noise}
\begin{figure}[h]
 \centering
 \includegraphics[scale=0.8]{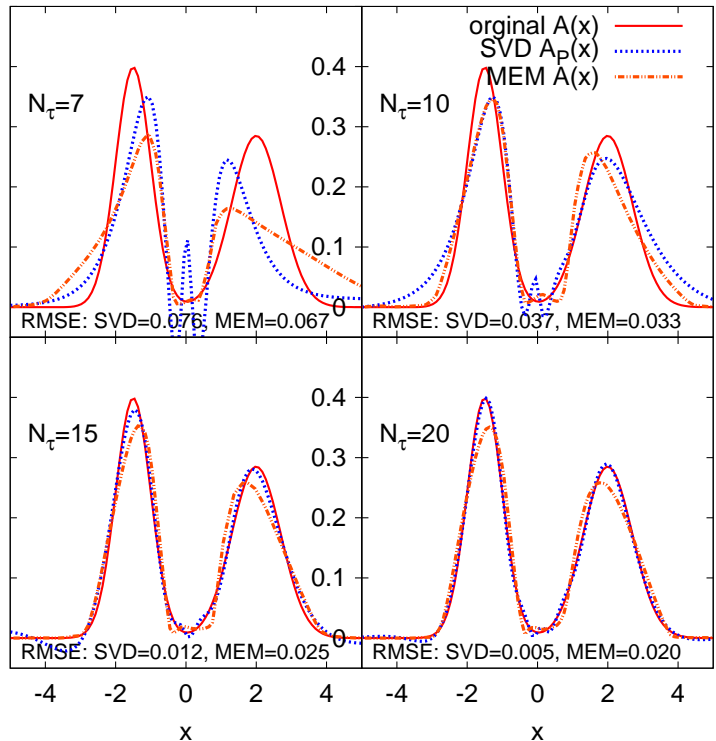}
 \caption{(Color online) The reconstruction of the artificial object function $A(x)$ by the SVD and MEM methods. The reconstruction is performed using $\Ntau$ uniformly spaced data points $G_{i}$ in the interval $[0,\beta=10]$
(noiseless data). A very good agreement between the normal solution and the original object 
is achieved if $\Ntau\geqslant 20$. The root mean square error (RMSE) 
for both methods is displayed at the bottom of the figure.}
 \label{fig:spectralTest1_A}
\end{figure}

As a first step the quality of reconstruction of the object as a function 
of the number of data points $\Ntau$ has been tested.
It was found that in order to reproduce the original object with a satisfactory accuracy, one has to use $\Ntau\geqslant 20$ data points; see \Fig{fig:spectralTest1_A}. Further increase in
the number of data points does not improve significantly the conformity between the solution and the object. It is related to the fact that the dimension of the subspace $\ASpace{M}$ increases 
linearly with an increase in the number of data points up to $\Ntau=20$ and then it saturates (for the presented example the dimension of $\ASpace{M}$-space is $20$ for $\Ntau=20$ and $25$ for $\Ntau=100$). 
Note also that the projected solution produced by the SVD method provides a very good approximation 
of the ``true'' object.

\subsection{Data with noise}
\begin{figure}[h]
 \centering
 \includegraphics[scale=0.8]{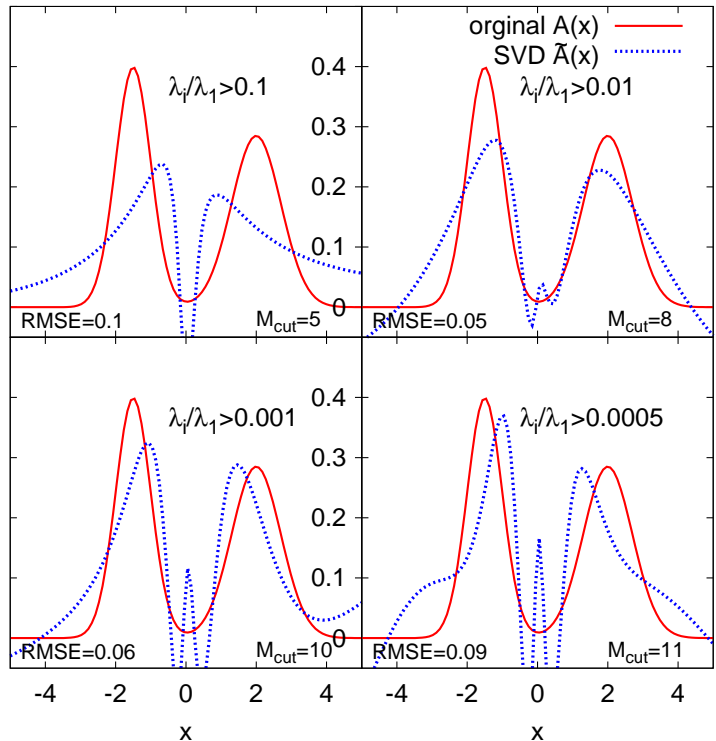}
 \caption{(Color online) Reconstruction of the artificial object function $A(x)$ by TSVD method. Reconstruction is performed using $\Ntau=25$ uniformly spaced data points $G_{i}$ in the range $[0,\beta=10]$, perturbed by Gaussian noise $\Delta G_{i}=\mathcal{N}(0,G_{i}/100)$. In the reconstruction procedure only those terms were included for which $\lambda_{i}/\lambda_{1}$ is larger than a given cut-off parameter. 
$\rcut$ denotes the number of the singular function included in TSVD expansion.}
 \label{fig:spectralTest2_As}
\end{figure}

In order to test the ability of reconstruction in the presence of noise
each value $G_{i}$ 
has been perturbed by the Gaussian noise of zero mean value and the standard deviation equal to 1\% of $G_{i}$. 
In the case of an SVD method the object 
has been reconstructed using \Eq{eqn:SVDA_tilde} for various cut-off parameters $\lambda_{i}/\lambda_{1}$ ($\lambda_{1}$ is the highest singular value); see \Fig{fig:spectralTest2_As}. Note that solutions with a cut-off parameter bigger than the relative error $0.01$ do not guarantee the reproduction of the imaginary time correlator within its error bars.
It is clearly seen that with a decreasing value of the cut-off parameter the quality of reconstruction 
increases, and the ``optimal'' cut-off parameter is $\lambda_{i}/\lambda_{1}\approx0.01$. 
Further decrease in the cut-off leads to the inclusion of an 
unjustified structure (strong fluctuations) into the shape of the reconstructed object.

\begin{figure}[h]
 \centering
 \includegraphics[scale=0.8]{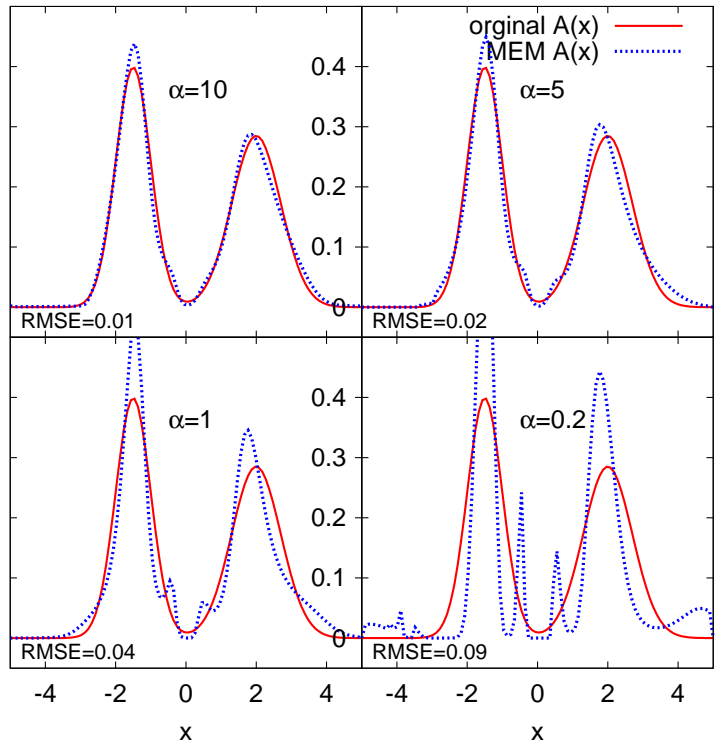}
 \caption{(Color online) Reconstruction of the artificial object function $A(x)$ by self-consistent MEM for different values of parameter $\alpha$.}
 \label{fig:spectralTest2_Am}
\end{figure}

In the case of MEM the quality of reconstruction is a function of $\alpha$ parameter; see \Fig{fig:spectralTest2_Am}. A class of assumed models has been chosen according to the prescription:
\beq
{\cal M}(x; c_{1},c_{2},\mu_{1},\mu_{2},\sigma_{1},\sigma_{2})=c_{1}N(x;\mu_{1},\sigma_{1}) + c_{2}N(x;\mu_{2},\sigma_{3}).
\label{eqn:doublegaussian}
\eeq

It was found that there exists a critical value of $\alpha$ parameter which separates ``smooth'' and 
``rigged'' solutions. It corresponds to the value which minimize the total MEM errors as discussed
in Ref. \cite{ghs}.
Moreover, it turns out that the ``self-consistent'' algorithm 
always converge to the same solution irrespective of initial values of parameters which define the class of models. 

\subsection{Impact of external constraints and \apriori information}
\begin{figure}[h]
 \centering
 \includegraphics[scale=0.8]{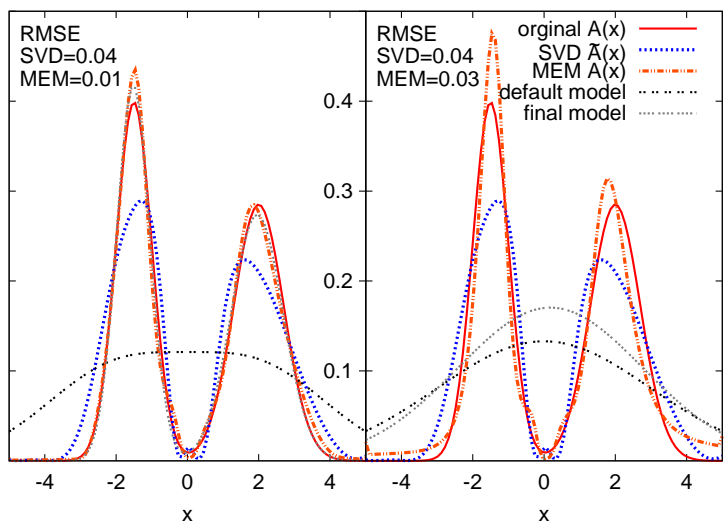}
 \caption{(Color online) The reconstruction ability of the spectral function for the full problem (data with noise + external constraints) of the SVD and MEM methods. The left panel shows the solution of the self-consistent MEM with  a combination of two Gaussians functions as a default model class. The right panel shows the solution of the self-consistent MEM with Gaussian functions as a default model class.}
 \label{fig:spectralTest3}
\end{figure}

In the following the impact of the external constraints on the reconstruction quality
has been tested for both MEM and SVD methods; see \Fig{fig:spectralTest3}. The external constraints influence strongly the SVD method. Note also that the solution produced by an SVD method 
is not an accurate reconstruction of the input spectral function. It is due to the fact 
that the solution produced by the SVD method is a projection of the ``true'' spectral function onto 
the ``visible'' subspace, where the problem is well posed. 
The main advantage of an SVD approach is that it does not require any \apriori information. 
Consequently the SVD solution can deliver very useful information concerning the default model or a class of default models for MEMs. 
In this particular test the SVD solution suggests that it is profitable to choose
the default model as a combination of two Gaussians (left panel), given by \Eq{eqn:doublegaussian}. 
One can also use the SVD solution as a default model for MEMs.

The tests presented above suggest that the maximum entropy method combined with ``self-consistent'' engine provides the most accurate solutions. Even if the class of the models is not correctly prepared, the self-consistent solution still well reproduces the input object. The right panel presents the case where the class of default model Gaussian functions $N(x;\mu,\sigma)$ was used.

Therefore the best methodology of producing the solution is suggested to be the following:
\begin{enumerate}
 \item Create an SVD solution and apply it to construct the class of default models ${\cal M}(x;\vec{f})$;
 \item Use the ``self-consistent'' MEM with constructed class of models ${\cal M}(x;\vec{f})$ to
       produce final solution.
\end{enumerate}

\subsection{Resolution limit}
In the case of physical applications where the object is associated with the spectral weight function, 
an extremely important question needs to be answered: is the spectral function unimodal or bimodal? 
It is well known that the distinct peaks of a bimodal spectral function may be overlooked 
during the reconstruction process if the distance between peaks is smaller than some critical value, which
defines the \textit{resolution limit}. 

To quantitatively estimate the resolution limit, the artificial object function consisting of two delta 
functions separated by $2\Delta_{0}$ distance have been considered. Namely, $A(x)=\delta(x+\Delta_{0})+\delta(x-\Delta_{0})$. For this function 
the imaginary time propagator has been generated for $\Ntau=25$ uniformly distributed data points in the
interval $[0,\beta]$, where now $\beta$ is treated as a parameter (in this case the inverse of $\beta$ has 
the physical meaning of temperature). The resolution limit $\Delta_{0}^{(\textrm{min})}$ is defined as a
minimal value of $\Delta_{0}$ for which the bimodal structure of the object can still be reconstructed and in general is a function of $\beta$.
\begin{figure}[t]
 \centering
 \includegraphics[scale=0.9]{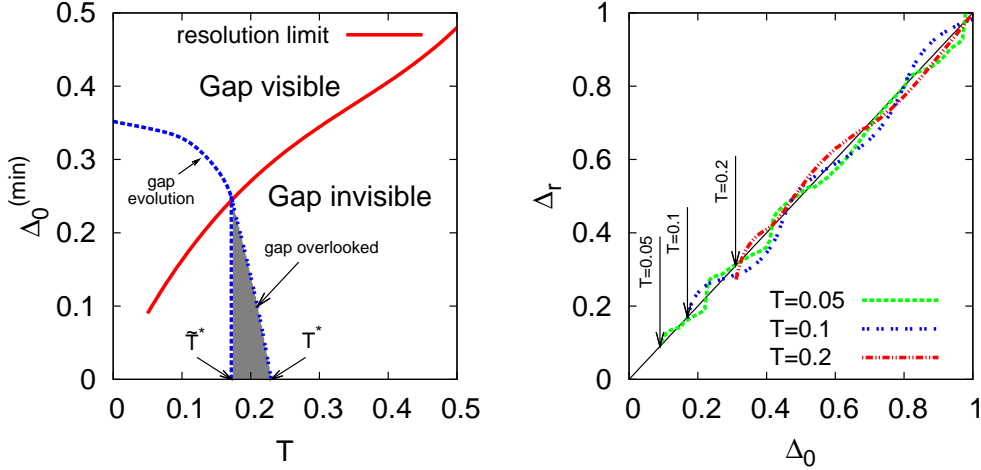}
 \caption{(Color online) Left panel: the resolution limit as a function of temperature $T=1/\beta$ (red solid line). Above the limit it is possible to reconstruct the gap (defined
as a distance between two peaks). The sketch of a typical evolution of the physical gap (dotted blue line)
is also plotted. The gap can be properly reconstructed up to $\Tmax$ temperature. 
 $T^{*}$ denotes the temperature for which the gap vanishes, assuming
that the reconstruction provide an exact object. 
Right panel: the value of the gap $\Delta_{\textrm{r}}$ reconstructed by the SVD method  versus 
the true value $\Delta_{0}$. The support function is $S(x,-2,2)$. Arrows indicate 
the minimal value of the gap (resolution limit), where the bimodal structure of the reconstructed spectral 
function appears.}
 \label{fig:spectralTest4}
\end{figure}

Results of the presented tests are shown in \Fig{fig:spectralTest4}. For both methods (SVD and MEM)
the existence of the finite resolution limit has been found. It increases with a decreasing temperature $T=1/\beta$ (left panel).
Within the presented approach it is possible to reconstruct the gap (defined
as a distance between two peaks) only for 
the temperatures for which it is larger than the reconstruction limit.

Let us consider a process of reconstructing the physical gap, which is a decreasing function of temperature and eventually vanishes at some temperature $T^{*}$.
At a certain temperature $\Tmax<T^{*}$ the gap becomes comparable with the reconstruction limit. 
Up to this temperature the reconstructed value of the gap $\Delta_{\textrm{r}}$ 
agrees very well with the true value $\Delta_{0}$ (right panel). At the temperature $\Tmax$ the value $\Delta_{\textrm{r}}$
drops to zero. It means that the methods provide in practice a lower bound for the temperature at which the true gap vanishes.

\section{Conclusions}
Library \verb|LINPRO| for solving arbitrary linear inverse problems with external constraints
has been presented. The library uses the Maximum Entropy Method and the SVD methods. An object-oriented implementation ensures that the package acquires a unified and user friendly interface.
As an example we have applied the new package to solve the typical problem of computational physics: analytic continuation of imaginary time propagator to real frequencies.

%%%%%%%%%%%%%%%%%%%%%%%%%%%%%%%%%%%%%%%%%%%%%%

\section{Acknowledgments}
The support from the Polish Ministry of Science under contract N N202 128439 and from the DOE under grants DE-FG02-97ER41014 and DE-FC02-07ER41457 is acknowledged. One of the authors (G.W.) acknowledges the Polish Ministry of Science for the support within the program ``Mobility Plus - I edition'' under contract No. 628/MOB/2011/0.
This work has also been partially supported by COMPSTAR, an ESF Research Networking Programme.
Calculations were performed at the Interdisciplinary Centre for Mathematical and Computational Modelling (ICM) at Warsaw University. 

%%%%%%%%%%%%%%%%%%%%%%%%%%%%%%%%%%%%%%%%%%%%%%%%%%%

\end{document}